\documentstyle[epsfig,subfigure]{mn}

\def\meanchi{\overline{\chi^2}}
\newcommand{\bi}[1]{\mbox{\boldmath $#1$}}
\title{Return mapping of phases and the analysis of the gravitational
clustering hierarchy}

\author[Chiang, Coles \& Naselsky] {Lung-Yih Chiang$^{1}$, Peter
Coles$^{2}$, and Pavel Naselsky$^{1,3}$\\
$^1$ Theoretical Astrophysics Center, Juliane Maries Vej 30,
DK-2100,  Copenhagen, Denmark \\
$^2$ School of Physics \& Astronomy, University of
Nottingham, University Park, Nottingham NG7 2RD, United Kingdom \\
$^3$ Rostov State University, Zorge 5, 344090 Rostov-Don, Russia}

\date{Accepted 2002 ???? ???; Received 2002 ???? ???}

\begin{document}
\maketitle

\begin{abstract}
In the standard paradigm for cosmological structure formation,
clustering develops from initially random-phase (Gaussian) density
fluctuations in the early Universe by a process of gravitational
instability. The later, non-linear stages of this process
involve Fourier mode-mode interactions that result in a complex pattern of
non-random phases. We present a novel mapping technique that
reveals mode coupling induced by this form of nonlinear interaction
and allows it to be quantified statistically. The phase mapping
technique circumvents the difficulty of the circular characteristic of
$\phi_{\bi k}$ and illustrates the statistical significance of phase
difference at the same time. This generalized method on phases allows
us to detect weak coupling of phases on any $\Delta{\bi k}$ scales.

\end{abstract}

\newcommand{\etal}{{et~al.~}}

\section{Introduction}

The morphology of the large-scale structure in the Universe is
that of a complex hierarchy of nodes, filaments and sheets
interlocking large voids. The Fourier-space description of such a
pattern is dominated by the properties of the phases rather than
the amplitudes of the Fourier modes \cite{c3}. According to the
prevailing theoretical ideas this pattern developed by a process
of gravitational instability from an amorphous pattern of density
fluctuations characterized by a Gaussian field with random phases.
Since the non-random phases of the present structure have grown
from random-phase initial perturbations then there is strong
motivation for understanding how phase information develops within
this paradigm and to construct a statistical description of galaxy
clustering that could be used as a test of the basic idea.

Unfortunately, quantifying the properties of Fourier phases is
difficult for a number of technical reasons, so their use in
statistical studies has so far been limited. Ryden \& Gramann
(1991), Soda \& Suto (1992) and Jain \& Bertschinger (1996)
focused on the evolution of individual phases away from their
initial values but since the initial phases are unknown these
studies can not be used as the basis of a statistical descriptor.
The pattern of association between phases is subtle and hard to
visualize which makes a statistical test hard to construct {\it a
priori}.

As the first step in a different approach towards quantifying
phase information, Coles \& Chiang (2000) proposed a colour
representation method to visualize phase coupling that at least
reveals qualitatively how phase information arises during the
evolution of $N$-body experiments but does not in itself
constitute a statistical descriptor. In a related study, Chiang \&
Coles (2000) quantified phase information using a statistic
derived from the Shannon entropy of the distribution of successive phase
differences. This study displayed interesting relationships
between phase entropy and gravitational clustering but still did
not provide a general statistical description.

In this paper we use a generalization of the concept of a return
map \cite{may,cc1} to transfer the phases of different Fourier
modes on to a bounded square upon which simple statistical tests
can be applied. In this way, we build upon the earlier studies
\cite{cc1,cc2} to construct a method that allows us to transform
the phase information in a clustering pattern into a more useful
form.

\section{Phase Coupling in the Nonlinear Regime}
The mathematical description of an inhomogeneous Universe revolves
around the dimensionless density contrast, $\delta({\bi x})$,
which is obtained from the spatially-varying matter density
$\rho({\bi x})$ via
\begin{equation} \delta ({\bi x}) =
\frac{\rho({\bi x})-\rho_0}{\rho_0},
\end{equation}
where $\rho_0$ is the global mean density. When the density
perturbation is small, the evolution of the density contrast can be
obtained analytically through {\it linear perturbation theory} from 3
coupled partial differential equations. They are the linearized continuity
equation,
\begin{equation}
{\partial\delta\over \partial t} = - {1\over a}{\bi \nabla_x}\cdot
{\bi v}, \label{eq:lCont}
\end{equation}
the linearized Euler equation
\begin{equation}
{\partial {\bi v}\over\partial t} + {\dot a\over a}{\bi v} = -
{1\over \rho a}{\bi \nabla_x} p -{1\over a}{\bi \nabla_x}\phi,
\label{eq:lEuler}
\end{equation}
and the linearized Poisson equation
\begin{equation}
{\bi \nabla_x}^2\phi = 4\pi G a^2\rho_0\delta. \label{eq:lPoisson}
\end{equation}

In these equations, $a$ is the expansion factor, $p$ is the pressure,
${\bi \nabla_x}$ denotes a derivative with respect to the comoving
coordinates ${\bi x}$, ${\bi v}=a \dot{\bi x}$ is the peculiar
velocity and  $\phi({\bi x},t)$ is the peculiar gravitational
potential. From Eq.~(\ref{eq:lCont})-(\ref{eq:lPoisson}), and if one
ignores pressure forces, it is easy to obtain an equation for the
evolution of $\delta$:
\begin{equation}
\ddot\delta + 2(\frac{\dot a}{a})\delta - 4 \pi G \rho_0 \delta = 0.
\label{eq:2ndorder}
\end{equation}
For a spatially flat universe dominated by pressureless matter,
$\rho_0(t) = 1/6\pi Gt^2$ and Eq.~(\ref{eq:2ndorder}) admits two
linearly independent power law solutions $\delta({\bi x},t) =
b_{\pm}(t)\delta_0({\bi x})$, where $\delta_0({\bi x})$ is the initial
condition, $b_+(t) \propto a(t) \propto
t^{2/3}$  is the growing  mode and $b_-(t) \propto t^{-1}$ is the
decaying mode.

It is useful to expand the density
contrast in Fourier series, in which $\delta$ is treated as a
superposition of plane waves:
\begin{equation}
\delta ({\bi x}) =\sum \tilde{\delta}({\bi k}) \exp(i{\bi k}\cdot
{\bi x}). \label{eq:fourier}
\end{equation}
The Fourier transform $\tilde{\delta}({\bi k})$ is complex and
therefore possesses both amplitude $|\tilde{\delta} ({\bi k})|$
and phase $\phi_{\bi k}$ where
\begin{equation}
\tilde{\delta}({\bi k})=|\tilde{\delta} ({\bi k})|\exp(i\phi_{\bi k}).
\label{eq:fourierex}
\end{equation}

In the standard picture of `gravitational instability' model for
the origin of cosmic structure, particularly those involving
inflation, the initial perturbations are Gaussian \cite{bbks}. The
most relevant property of Gaussian random fields is that they
possess Fourier modes whose real and imaginary parts are
independently distributed. In other words, they have phase angles
$\phi_{\bi k}$ that are independently distributed and uniformly
random on the interval $[0,2\pi]$. Terms in the perturbative
evolution equations for the Fourier modes that represent coupling
between different waves are of second (or higher) order in
$\delta$ and these are neglected in linear perturbation theory.
When fluctuations are small, i.e., during the linear regime, the
Fourier modes evolve independently (Eq.(~\ref{eq:2ndorder}) and
(\ref{eq:fourier})) and the Gaussian character is retained in the
linear regime, where the phases remain independent and uniformly
random. In the later stages of evolution, however, modes begin to
couple together. In this non-linear regime that Fourier phases
become non-random. For a thorough review of the theory and
implications of non-linear evolution from the point of view of
perturbation theory, see Bernardeau et al. (2002).

\begin{figure}
\centering
\epsfig{file=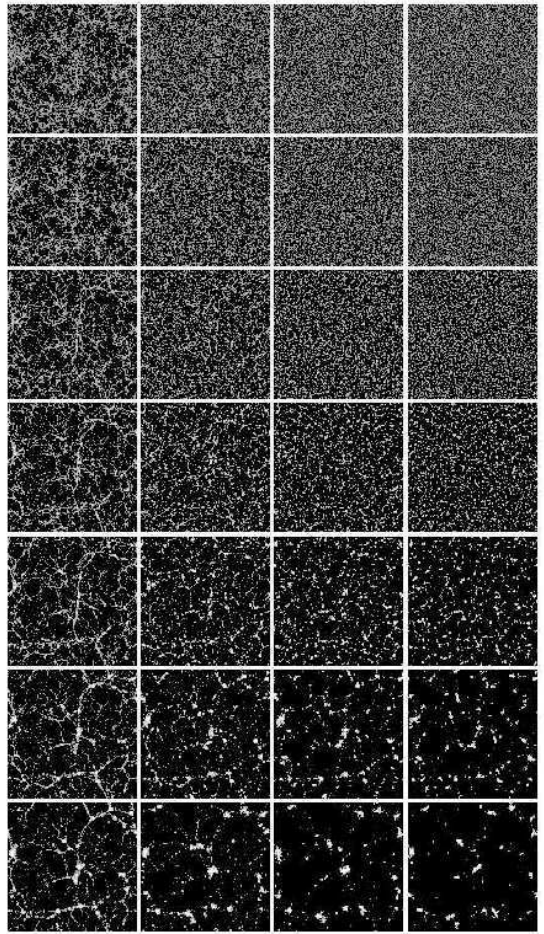,width=10cm}
\caption{2D $N$-body simulations for initial power spectral index,
from left to right, $n=-1$, 0, 1 and 2, respectively. We choose 7
stages, named stage 1 to 7 from top to bottom, with an increasing
level of non-linearity as described in the text. We control the
simulations by setting the same initial random phase
configuration for each index. Consequently, it is easy to see that the
evolving structures have density concentration at the same locations; the
difference is due to the initial spectral index: large $n$ will
produce more clumps, small $n$ will have filaments.}
\label{simulation}
\end{figure}

Standard methods of analysis proceed via the power-spectrum,
$P(k)$, essentially proportional to $|\tilde{\delta}({\bi k})|^2$.
The probabilistic properties of Gaussian random fields are
completely specified by knowledge of $P(k)$. Higher-order
quantities based on $\tilde{\delta}({\bi k})$ can also be defined,
such as the bispectrum \cite{peebles,mat,s98,s99,v20,v21,v22},
which vanishes for  Gaussian fields, or quantities related to
 correlations of $|\tilde{\delta}({\bi k})|^2$
\cite{stirling}. Phase coupling results in a non-Gaussian field in
which the bispectrum and higher-order polyspectra may be non-zero
\cite{wc}. Phase information is at the heart of non-linear galaxy
clustering.

\begin{figure*}
\centering
\epsfxsize 8cm
\subfigure[]{\epsfbox{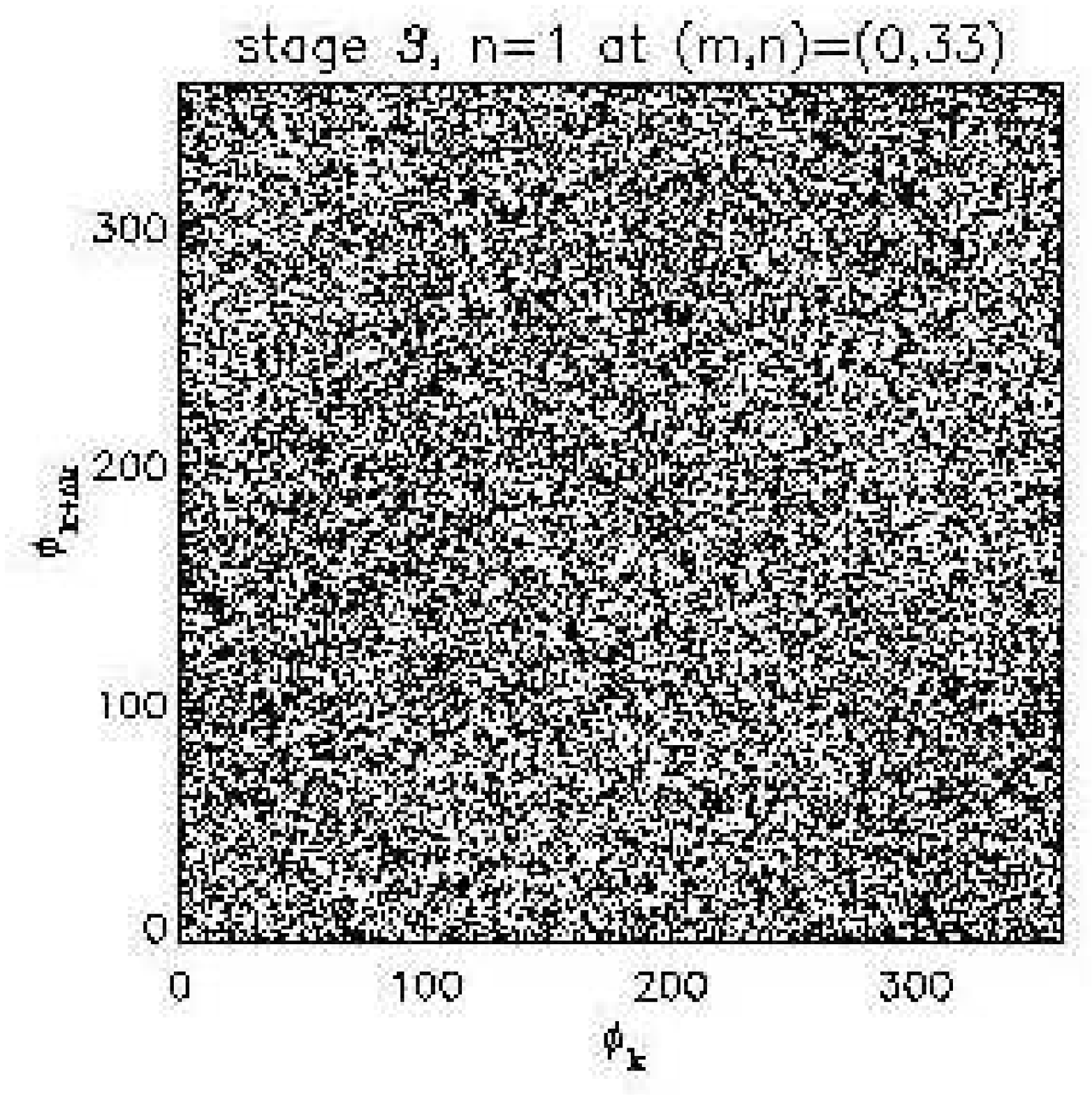}}
\subfigure[]{\epsfbox{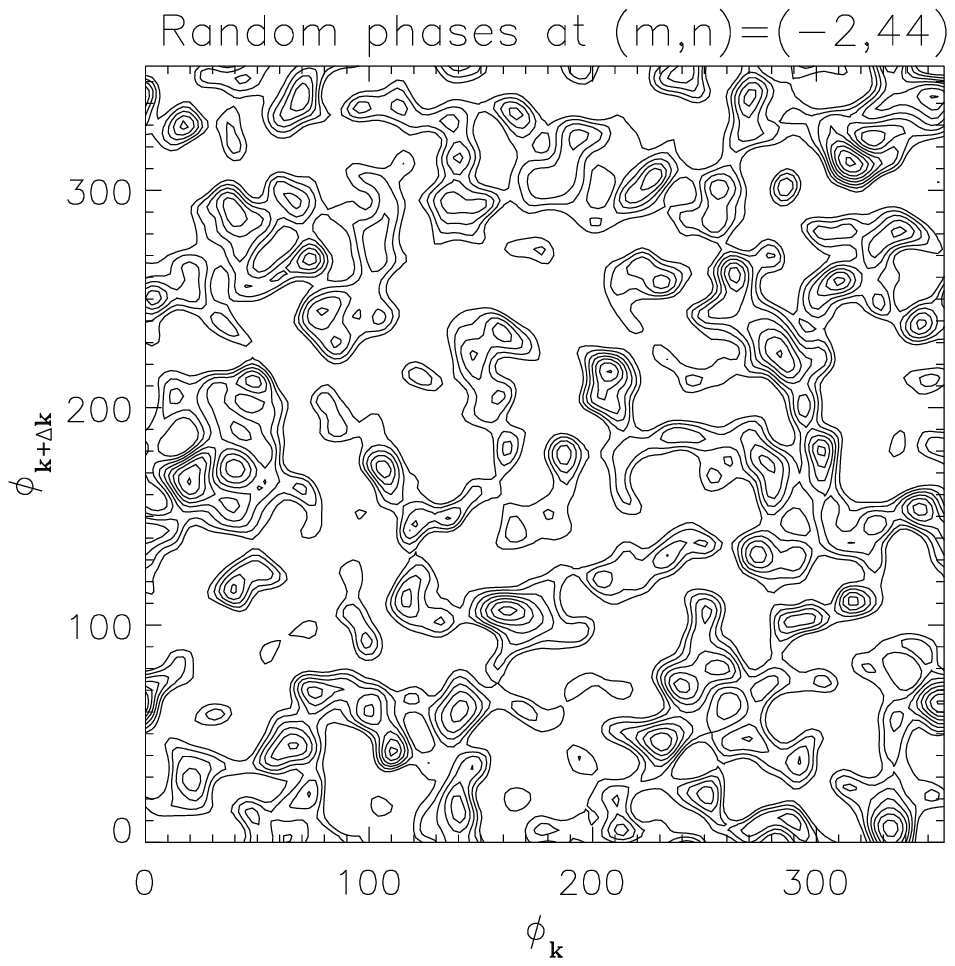}}\\
\subfigure[]{\epsfbox{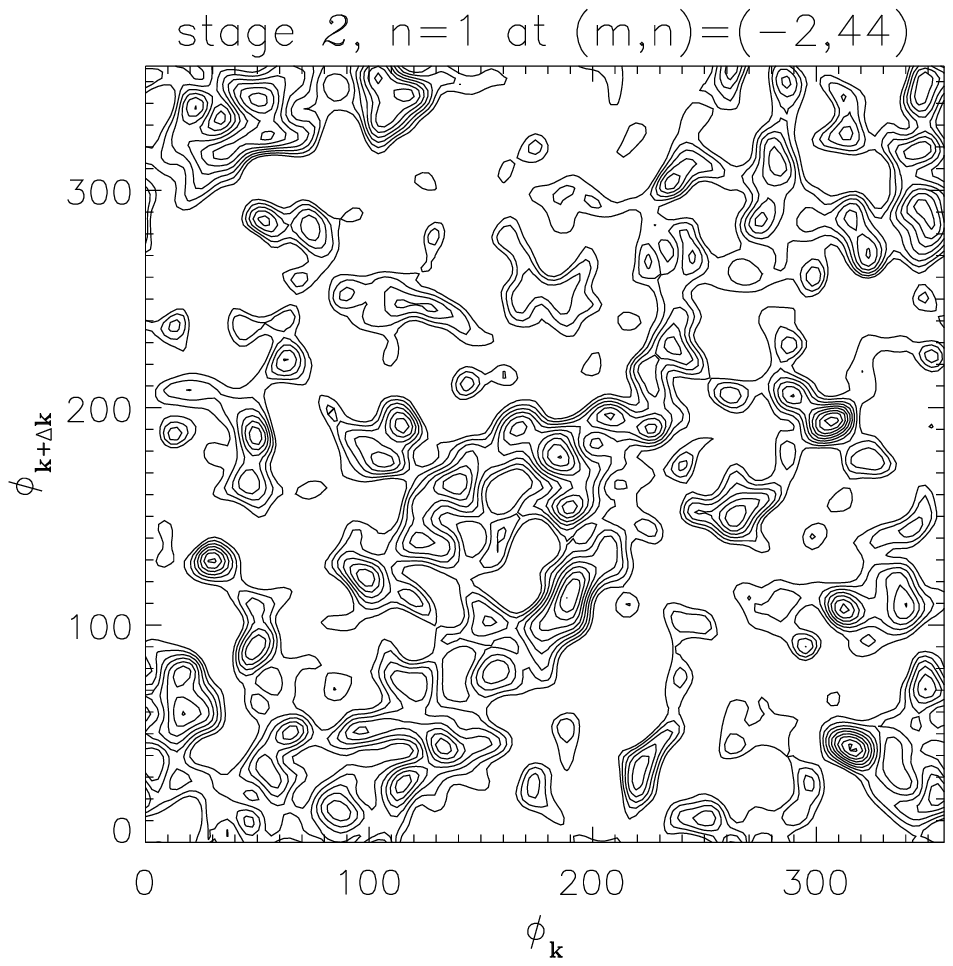}}
\subfigure[]{\epsfbox{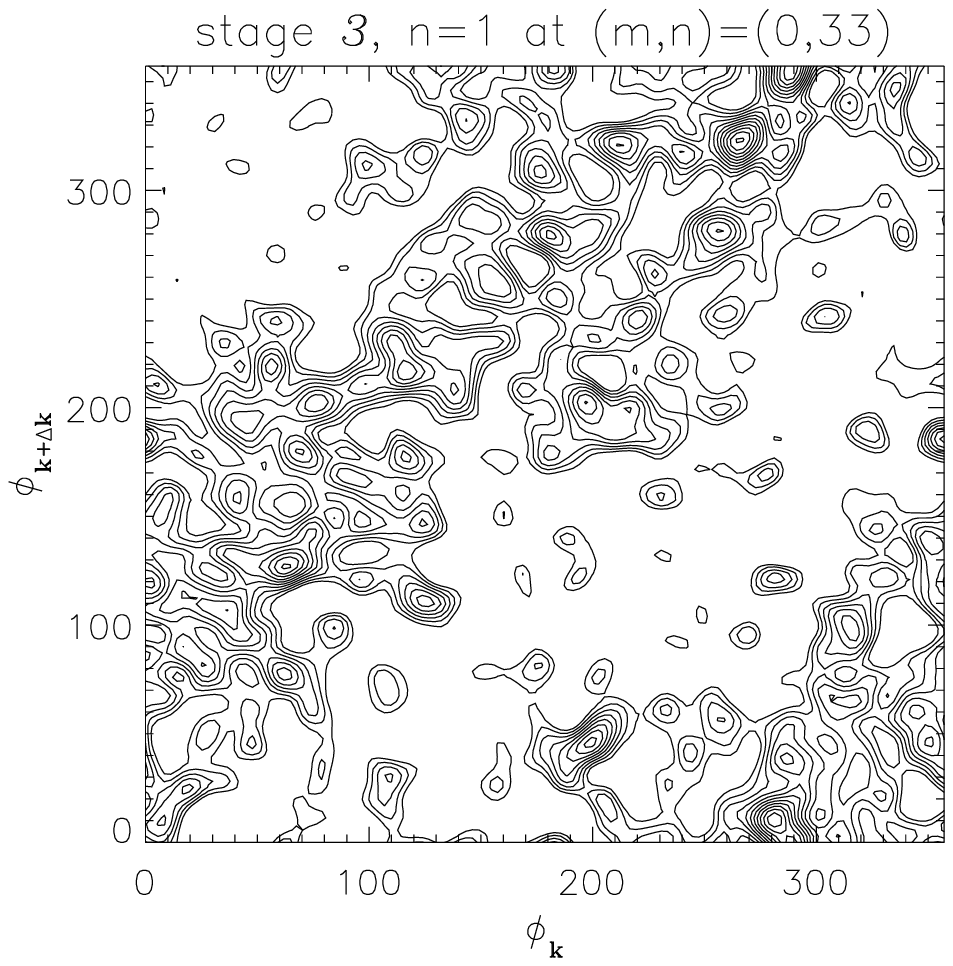}}\\
\caption{Phase mapping onto the return map. The horizontal and
vertical axes represent $\phi_{\bi k}$ and $\phi_{{\bi k}+\Delta{\bi
k}}$, ranging from $0^\circ$ to $360^\circ$. In panel (a) the phases
are taken from the stage 3, spectral index $n=1$ of the $N$-body
simulations shown in Fig.~\ref{simulation} and the scale of phase
mapping is $\Delta {\bi k} \equiv (m,n)=(0,33)$. The phases are
condensed along the diagonal strip. Panels (b), (c) and (d) are the
smoothed versions of various phase maps where the contour levels are
drawn upwards starting from the mean value. Panel (b) is from a
random-phase realization and (c) is that of stage 2, $n=1$. For
comparison, both (b) and (c) are mapping from the same $\Delta{\bi k}$
scale $(m,n)=(-2,44)$. Panel (d) is the smoothed version of panel
(a).} \label{mapping}
\end{figure*}

\section{Directional Phase Mapping}

There are two principal difficulties involved in constructing a
statistic from Fourier phases. One is that because phases reflect
the morphology, their values change according to the position of
structural features \cite{c3}. For example, the phases of a peak
in the form of Dirac-$\delta$ function $\delta_D(x-x_0)$,
$\phi_k=kx_0$ suffer change in slope along the $k$-axis when there
is shift of the peak $\delta_D(x-x_0-x^{'})$, the phases being
$\phi_k=k(x_0+x^{'})$. If a pattern is statistically homogeneous,
any descriptor of it should be translation-invariant and this is
manifestly not true of the phases themselves. The other problem is
that phases are of circular measures and therefore defined modulo
$2\pi$. Traditional measures of association, such as covariances
of the form $\langle \phi_i \phi_j \rangle$, are based on the
assumption that the measure associated with the variable is linear
and are therefore not appropriate to cases where values separated
by $2\pi$ are in fact equal in value.

To address these problems, Chiang \& Coles (2000) used the phase
difference (or phase gradient), $D_k$, defined in one dimension by
\begin{equation}
D_k\equiv\phi_{k+1}-\phi_{k},
\end{equation}
i.e. for neighbouring phases. In two or
three dimensions differences can be taken in orthogonal
directions. The quantity $D_k$ has the twin advantages that for random
phases it is also random but upon translation $x'$ it changes by a
constant $x'$ for all $k$. The statistical properties of the set of differences
$D_k$ contain information about the correlations of neighbouring
Fourier modes. Strong correlation of the neighbouring modes at large
$k$ corresponds to the highest peak in the clustering
pattern. Naselsky, Novikov and Silk (2002) used this characteristic to
extract point sources in the CMB map. Moreover, Chiang et al. (2001)
used the phase analysis for extracting the in-flight beam shape
properties of CMB experiments. To construct a more general
description of phase coupling we need to extend this method to modes
that are not necessarily neighbours. We do this by constructing a {\it
directional phase map}, based on the return maps used in non-linear
dynamics \cite{may}.

The basic idea is simple and based on a study of one-dimensional
examples contained in Chiang \& Coles (2000) which provides a
useful illustration of the more general approach. With a set of phases
$\phi_k$ from the Fourier transform of a one-dimensional process, one
can plot a map of $\phi_k$ against $\phi_{k+1}$ for each pair
$(\phi_k, \phi_{k+1})$. If the phases are random this will be a
scatter plot with points distributed randomly within the bounded
square of side $[0,2\pi]$. If there is association between
neighbouring phases the plot will contain correlations; the quantity
$D_k$ is sensitive to linear association. If the spatial pattern
consists of a single high-amplitude peak the points display linear
association and are mapped into a diagonal lines on the diagram.

\begin{figure*}
\centering
\epsfxsize 9cm
\subfigure[]{\epsfbox{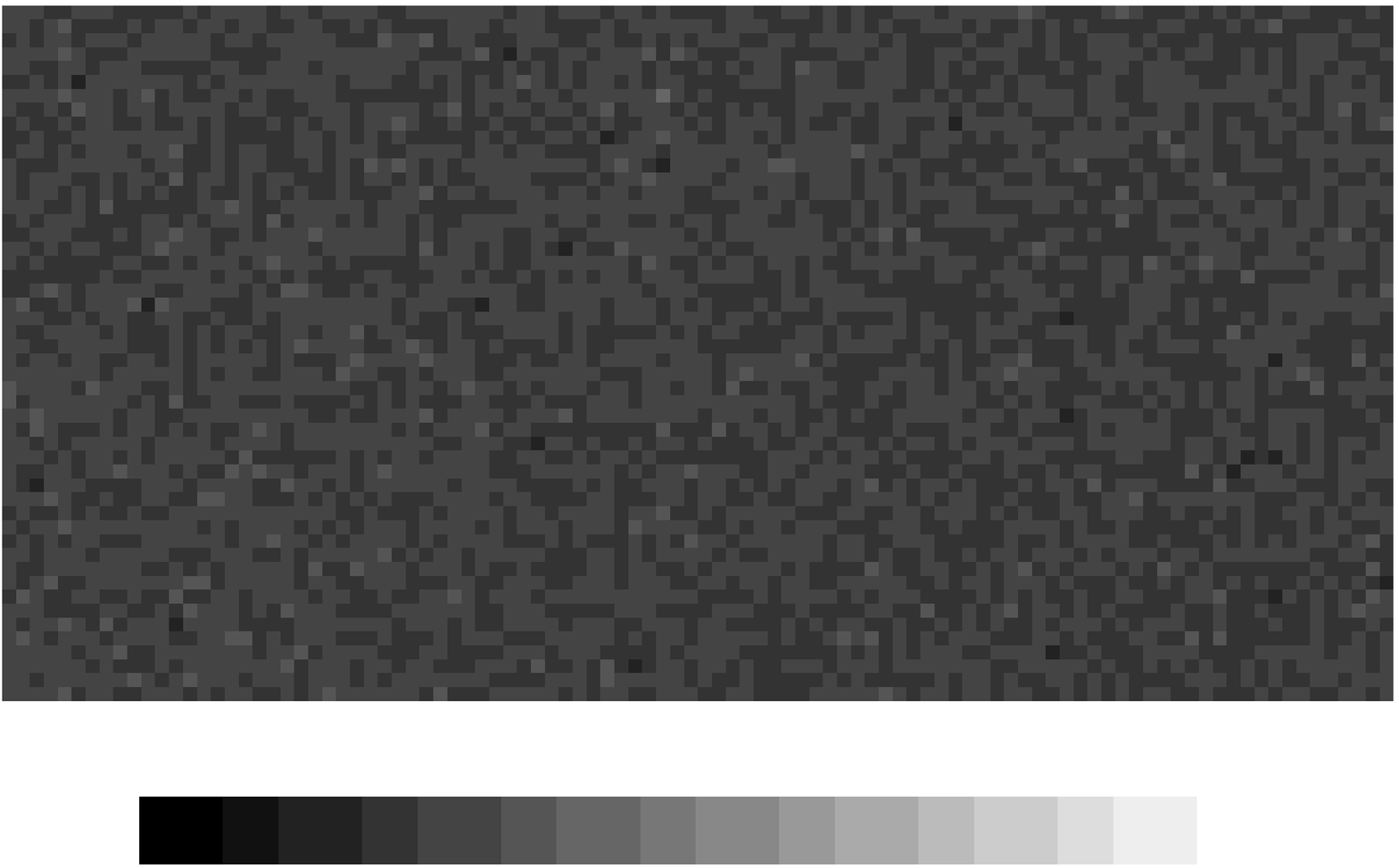}}
\put(-18.,20.){$50$}
\put(-255.,28){$1$}
\put(-132.,20.){$m$}
\put(-255.,20.){$-49$}
\put(-42.,-8.){$0.05$}
\put(-230.,-8.){$0.01$}
\put(-260,140){$50$}
\put(-257,85){$n$}
\put(-145.,152){stage 2, n=1}
\put(-224,1){\framebox(190,11)[t]}
\epsfxsize 9cm
\subfigure[]{\epsfbox{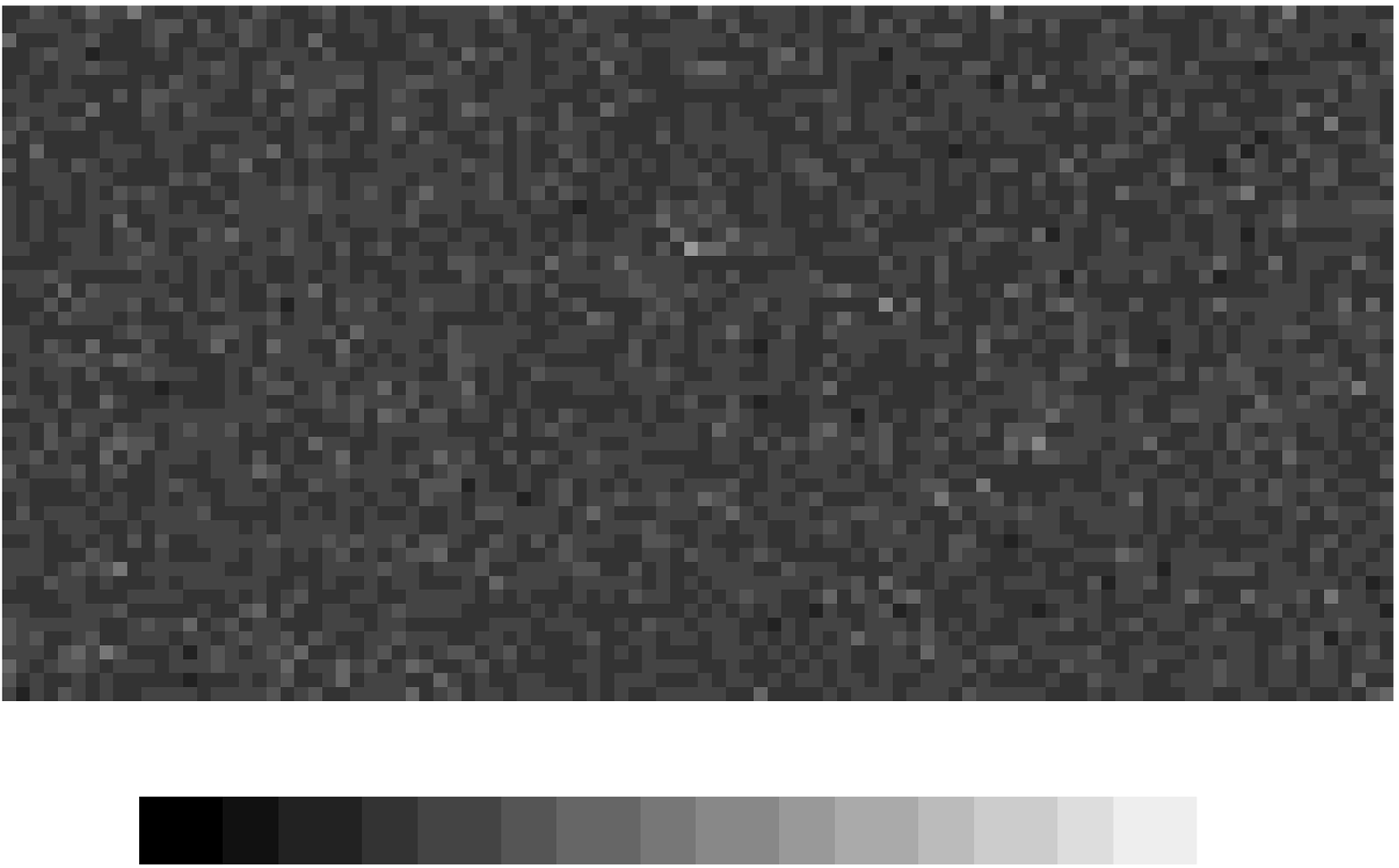}}
\put(-18.,20.){$50$}
\put(-255.,28){$1$}
\put(-132.,20.){$m$}
\put(-255.,20.){$-49$}
\put(-42.,-8.){$0.05$}
\put(-230.,-8.){$0.01$}
\put(-260,140){$50$}
\put(-257,85){$n$}
\put(-145.,152){stage 3, n=1}
\put(-224,1){\framebox(190,11)[t]} \\
\epsfxsize 9cm
\subfigure[]{\epsfbox{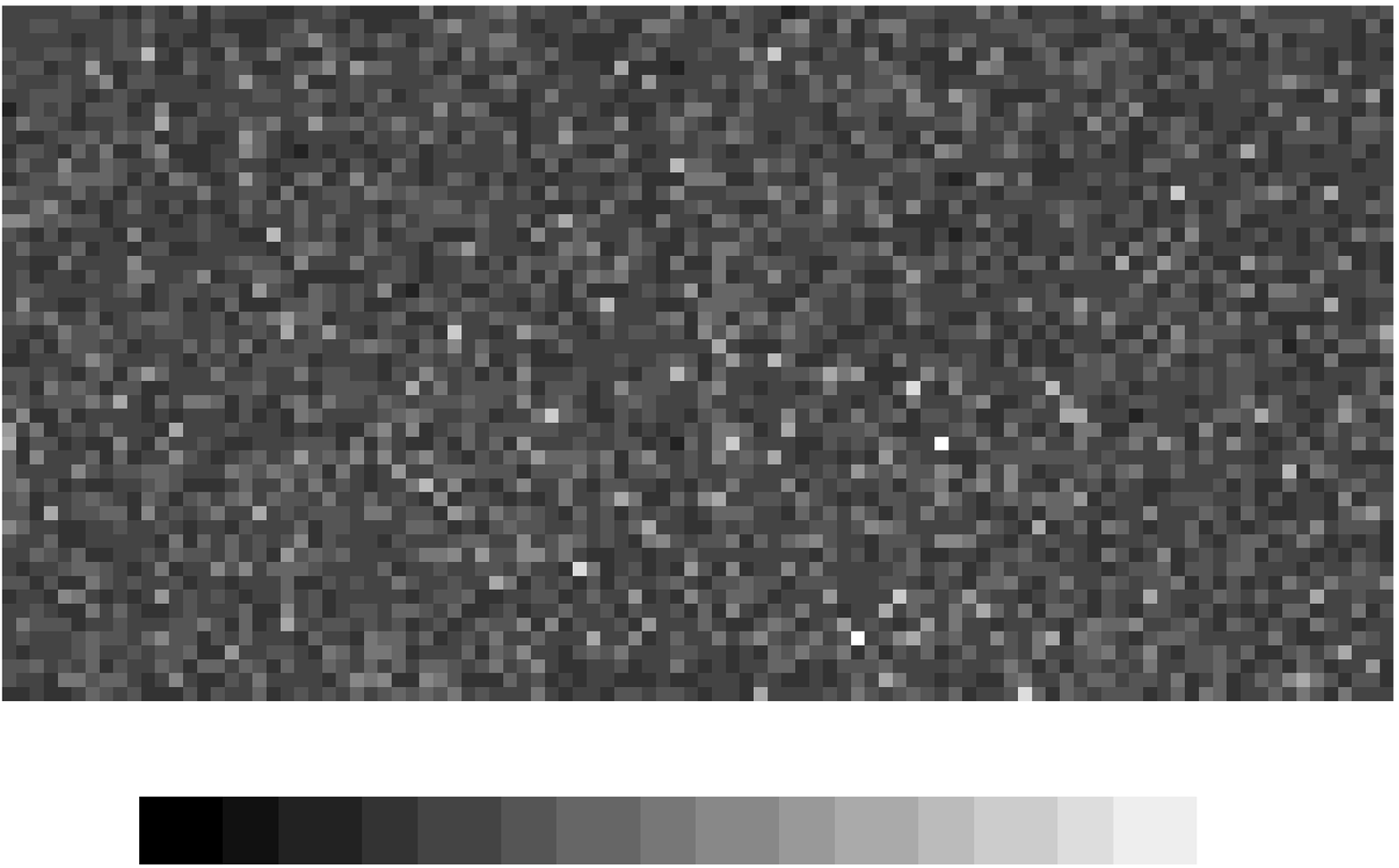}}
\put(-18.,20.){$50$}
\put(-255.,28){$1$}
\put(-132.,20.){$m$}
\put(-255.,20.){$-49$}
\put(-42.,-8.){$0.05$}
\put(-230.,-8.){$0.01$}
\put(-260,140){$50$}
\put(-257,85){$n$}
\put(-145.,152){stage 4, n=1}
\put(-224,1){\framebox(190,11)[t]}
\epsfxsize 9cm
\subfigure[]{\epsfbox{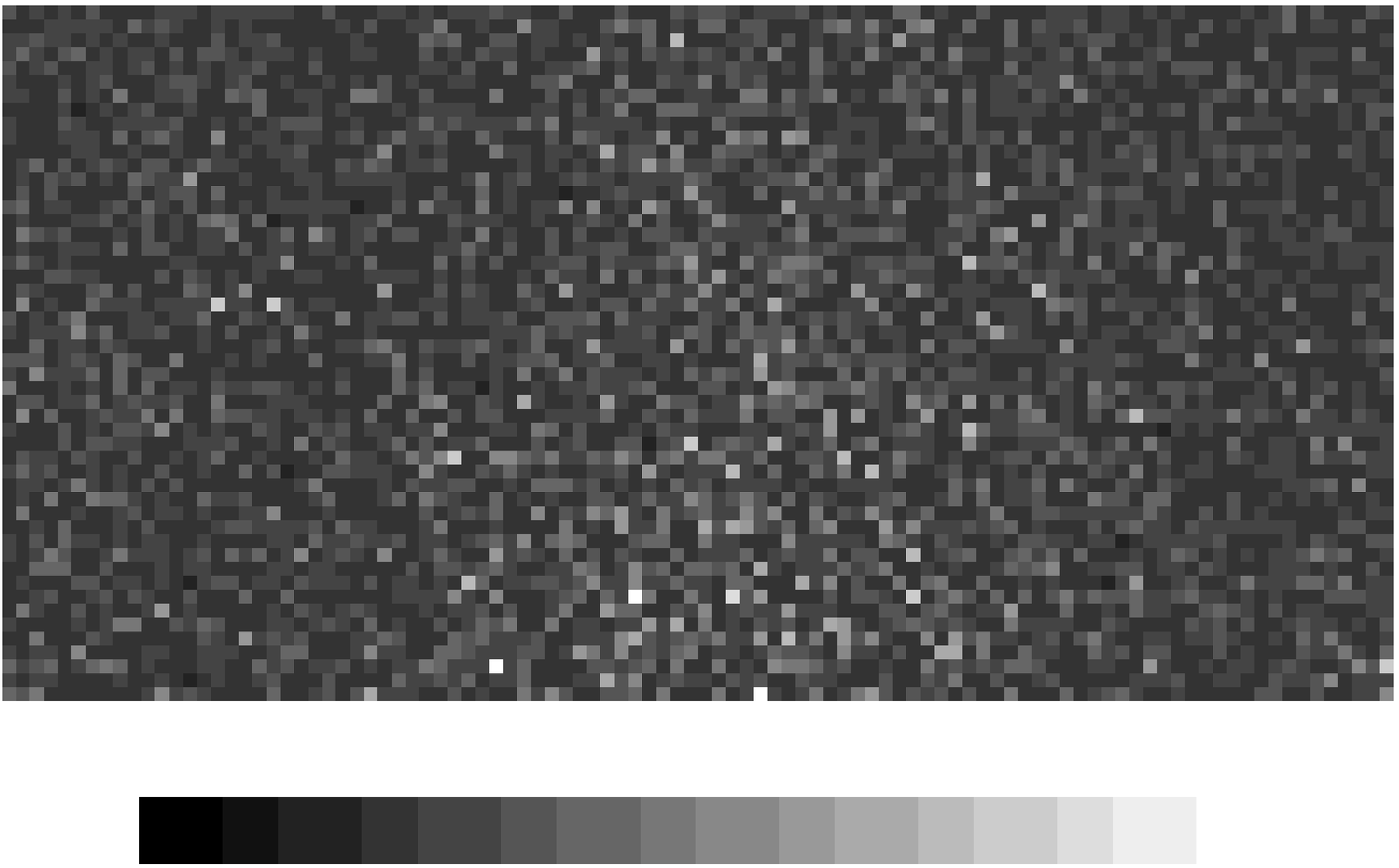}}
\put(-18.,20.){$50$}
\put(-255.,28){$1$}
\put(-132.,20.){$m$}
\put(-255.,20.){$-49$}
\put(-42.,-8.){$0.1$}
\put(-230.,-8.){$0.0$}
\put(-260,140){$50$}
\put(-257,85){$n$}
\put(-145.,152){stage 5, n=1}
\put(-224,1){\framebox(190,11)[t]}
\caption{The $\meanchi$ statistics on grey scale for different
realisations on the $(m,n)$ plane. The 4 panels correspond to the
4 realisations of Fig.~\ref{simulation}, stage 2, 3, 4 and
5, spectral index $n=1$ of the 2D $N$-body simulations. In order to
show clearly the maximal points, we set the same grey scale for panel
(a), (b) and (c), and note that the grey scale of (d) is
different. The maximal $\overline{\chi^2}$ for stage 2 is $2.53 \times
10^{-2}$ at $(m,n)=(-2,44)$, and $3.27 \times 10^{-2}$ for stage 3 at
$(m,n)=(0,33)$, both mapping of the particular scale being shown
in Fig.~\ref{mapping} (c) and (d). For panel (c) and (d) the
$\meanchi_{\max}=4.90 \times 10^{-2}$ and $1.17 \times 10^{-1}$ at
$(m,n)=(12,5)$ and $(-4,8)$, respectively, which indicates the scale
of phase coupling in the highly non-linear regime is at small $\Delta
{\bi k}$.}
\label{mnplane1}
\end{figure*}

In what follows, for illustration, we shall use two-dimensional
examples based on numerical simulations with periodic boundaries,
so we take $\Delta{\bi k} \equiv (m,n)$ where $m$ and $n$ are integers.
The simulations are done on a $512^2$ grid. In
Fig.~\ref{simulation} we show 4 sets of such $N$--body
simulations for initial power spectral index $n=-1$, 0, 1 and 2 (see Chiang \&
Coles 2000 for details of the simulations). The evolutionary
stages are characterized by an increasing scale of non-linearity
defined by $\left<\left(\delta\rho/\rho\right)^{2}\right>_{k_{\rm NL}}=
b^{2}_{+}(t) \int^{k_{\rm NL}}_{0}P(k)\: d^{2}k = 1$, where $b_{+}(t)$
is the growing mode of the linear density contrast and $P(k)$ is
the linear extrapolation of the initial power spectrum. This
definition of $k_{\rm NL}$ identifies the corresponding scale $2\pi
k_{\rm NL}^{-1}$ as the boundary between linearity and non-linearity.
The stages in Fig.~\ref{simulation} are chosen such that the scales of
non-linearity $k_{\rm NL}$ between any two successive stages vary by a
factor of 2. The levels of non-linearity of the stages are thus
$k_{\rm NL}=256$, 128, 64, 32, 16, 8 and 4.\footnote{To avoid
confusion with the panels in the captions, we re-name the stages as
1-7, which are originally named as stage $a$-$g$ in Chiang \& Coles (2000).}

We can use these simulations to illustrate how we extend the
return mapping between neighbouring phases ($\Delta {\bi k}=1$) to
pairs of phases with {\it any} $\Delta{\bi k}$ scales in $k$-space. We
map all pairs of phases $\phi(i,j)$ and $\phi(i+m,j+n)$ onto the $x$
and $y$ values of the return map. The axes therefore range over $[0, 2\pi]$ for both
$\phi({\bi k})$ ($x$) and $\phi({\bi k}+\Delta {\bi k})$ ($y$)
axes. For example, for $(m,n)=(4,6)$ we have points on the return
map $(\phi(i,j),\phi(i+4,j+6))$ for all $i\in [-255,256]$, $j
\in[1,256]$, i.e., all points $(\phi(1,1), \phi(5,7))$,
$(\phi(2,1),\phi(6,7))$, $(\phi(1,2),\phi(5,8)) \ldots$, from a 2D
Fourier transform of a realisation. This represents the
directional phase coupling for coupling scale $\Delta {\bi k} \equiv
(\Delta k_{x}, \Delta k_{y}) \equiv (m,n)$ in ${\bi k}$-space. The
neighbouring phase differences in the $k_x$-direction and
$k_y$-direction used by Chiang \& Coles (2000) simply corresponds to
$(m,n)=(1,0)$ and $(0,1)$ respectively.

In Fig.~\ref{mapping} (a) we show one example of phase mapping from
the realization of stage 3, spectral index $n=1$ simulation. The
particular $\Delta{\bi k}$ in this example is $(m,n)=(0,33)$. This
panel demonstrates how weak coupling between phase pairs with fixed
scale $\Delta{\bi k}$ can manifest itself in the return map as
non-uniform density in the map plane. 

This directional phase mapping approach circumvents the problem of
the circular character of $\phi_{\bi k}$ but does not attempt to
condense all the related information into a single quantity. It, on
the other hand, exploits all the information between all Fourier modes. For
example, the phase coupling of a 1D distribution can be expressed in a
(2D) return map. It therefore allows us to build simple statistics to test the
significance of general non-randomness. The phase difference
between any pair at a fixed scale becomes a single point on the
return map for that scale. The circular characteristic of $\phi_{\bi k}$
is transferred to a bounded square, topologically equivalent to a
torus owing to the periodicity of $x$ and $y$ axes. The bands seen
in Fig.~\ref{mapping} (a) therefore correspond to twisted linear features on
this torus.

The key advantage of directional phase mapping is that, for a
Gaussian random field, any directional phase mapping for {\it any} scale
$\Delta {\bi k} \equiv (m,n)$ should produce a random Poisson
distribution. Weak phase coupling will produce correlations at
large vectors $(m,n)$ while strong non-linearity will produce
highly non-uniform phase maps at all scales $(m,n)$.

\section{A $\chi^2$ Test on Phase Maps}

Once we have transferred the phase information onto a phase map
like that shown in Fig.~\ref{mapping} (a), many different
statistical tests can be used to analyse its properties. Here we
outline a simple yet powerful method.

First we smooth the return map. Smoothing enhances our
visualization of the pattern of phase coupling. In
Fig.~\ref{mapping} (b), (c), and (d) we show the contour maps of
the smoothed return map.  In these simple illustrative experiments
we divide the square of the return map into $128^2$ pixels, and we
bin the 32\,768 points so that, for a perfectly even distribution in
the map plane the occupation of each pixel is $\langle p(i,j)
\rangle =2$. Then we smooth this $128^2$ mesh by
\begin{equation}
p({\bi x},R)= (\sqrt{2\pi}R)^{-2} \int d^{2}{\bi x^{'}} \: p({\bi
 x^{'}}) \: \exp\left(-\frac{|{\bi x}-{\bi x^{'}}|^{2}}{2R^{2}}\right).
\label{eq:window}
\end{equation}

The contour levels are drawn upwards starting from the mean value.
Panel (b) is the contour of a smoothed return map from a
realization of random phases. Panel (c) is that from
stage 2, $n=1$ of Fig.~\ref{simulation} with $(m,n)=(-2,44)$. For
comparison, the coupling scale $\Delta {\bi k}$ for both (b) and (c) is set the
same. We can see that even for the mildly non-linear regime
represented by stage 2 (for which $\delta \simeq 1.8$), phase mapping
does reveal the existence of coupling by starting to condense on the
diagonal strip. Panel (d) is the smoothed version of (a) with
$(m,n)=(0,33)$, in which the condensed strip from (a) is much clearer.

We define a {\it mean} $\chi^2$ statistic as
\begin{equation}
\meanchi=\frac{1}{M}\sum_{i,j}\frac{\big[ p(i,j)-\overline
p\big]^2}{\overline p} \label{eq:chisquare}
\end{equation}
where $M$ is the number of pixels we assign on the return map and $\overline
p$ is the mean value for each pixel.

In Fig.~\ref{mapping}, the smoothing scale on the $128^2$ mesh is
$R=2$ and the contour levels are drawn starting from the mean
value. We use this pixel size and smoothing scale in all our
calculations of $\meanchi$ in the following, but one can vary the
scale as part of a statistical test.

\section{Simulation Results}
We now illustrate the results of this analysis using the 2D
$N$-body simulations described earlier. First we Fourier transform
the realizations of the $N^2$ mesh ($N=512$ in our simulations).
Because of the reality of the original distribution, and the
consequent Hermitian conjugate relations in the Fourier image,
only half of the Fourier transform contains independent
information. We end up with $N^2/2$ Fourier modes available, so we
take $-k_{N/2}+1$ to $k_{N/2}$ in $k_x$ axis and from 1 to
$k_{N/2}$ in $k_y$ axis.

\begin{figure}
\centering \epsfig{file=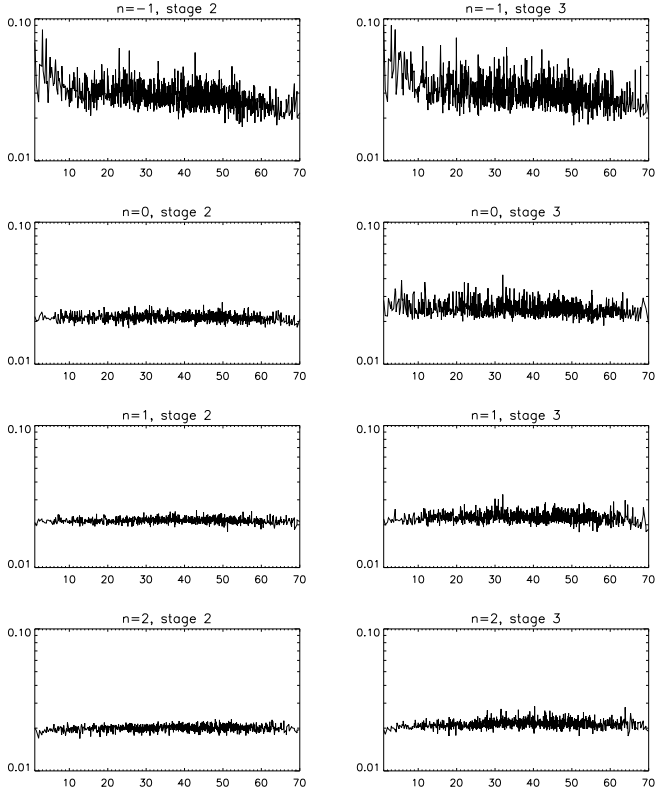,width=9cm} \caption{The
maximal $\meanchi$ is plotted against 1D
Fourier scale $|\Delta{\bi k}| \equiv \sqrt{m^2+n^2}$ for stage 2 and 3 of
simulations from initial spectral index $n=-1$, 0, 1 and 2. The
maxima are taken from each $|\Delta{\bi k}|$ ring of the $(m,n)$ plane. }
\label{maxchi23}
\end{figure}

We carry out a phase mapping for each $(m,n)$. The directional
phase mapping is performed for the vectors $(m,n)$, where
$m \in [-49,50]$ and $n \in [1,50]$ for the phase base
$\phi(i,j)$, where $i \in [-127,128]$, and $j \in [1,128]$, a quarter
of the available phases. The limited range of mapping vectors is chosen to
ensure that the map can be constructed without running out of sensible
wavenumbers.  Thus, in such a case, for each return map of $(m,n)$
there are $2 \times 128^2$ points. We set up $M=128^2$ pixels in this
$[0, 2\pi]$ square and then smooth the return map to decrease the
fluctuation. The $\meanchi(m,n)$ for each fixed scale $(m,n)$ is
calculated by Eq.~(\ref{eq:chisquare}).

Figure~\ref{mnplane1} shows  `supermaps' of all the phase
information contained in the $\meanchi(m,n)$ using a grey scale in
the $(m,n)$ plane from various realisations of simulations. We are
interested specifically on the mild non-linear regime, so in
Fig.~\ref{mnplane1} the panels (a), (b), (c)  and (d) correspond
to stages 2, 3, 4 and 5 of $n=1$ in Fig.~\ref{simulation},
respectively. The $m$-axis ranges from $-49$ to $50$ and $n$-axis
ranges from 1 to 50. Each pixel therefore displays the level of
phase coupling on a certain scale $(\Delta k_x, \Delta k_y) \equiv
(m,n)$ in terms of the $\meanchi$ statistics. The bright points in
Figure~\ref{mnplane1}  are direct indications of phase coupling
for the corresponding $\Delta {\bi k}$ scales.

\begin{figure}
\centering \epsfig{file=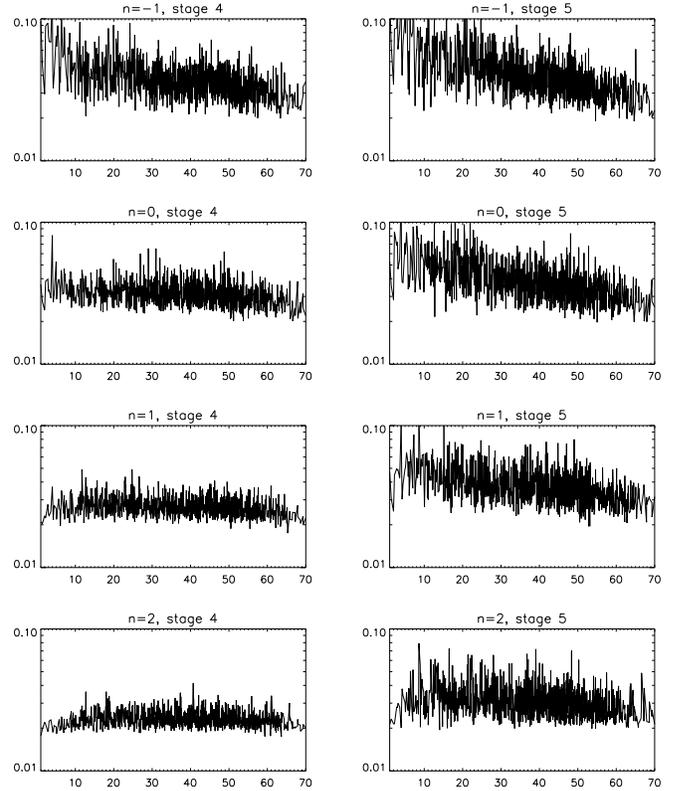,width=9cm} \caption{The
maximal $\meanchi$ is plotted against 1D
Fourier scales $|\Delta{\bi k}| \equiv \sqrt{m^2+n^2}$ for stage 4 and 5 of
simulations from initial spectral index $n=-1$, 0, 1 and 2. The
maxima are taken from each $|\Delta{\bi k}|$ ring of the $(m,n)$ plane.}
\label{maxchi45}
\end{figure}

The reason we present the specific scales $(m,n)=(-2,44)$ for
stage 2 and $(m,n)=(0,33)$ for stage 3, $n=1$ in Fig.~\ref{mapping}
is that the calculation of the $\meanchi$ statistics  shows them
to be `hotspots' of phase correlation. The scales of phase coupling for
those two panels correspond to the brightest points, i.e. the maximum
$\meanchi$ on the corresponding $(m,n)$ planes in Fig.~\ref{mnplane1}.

\begin{figure}
\centering \epsfig{file=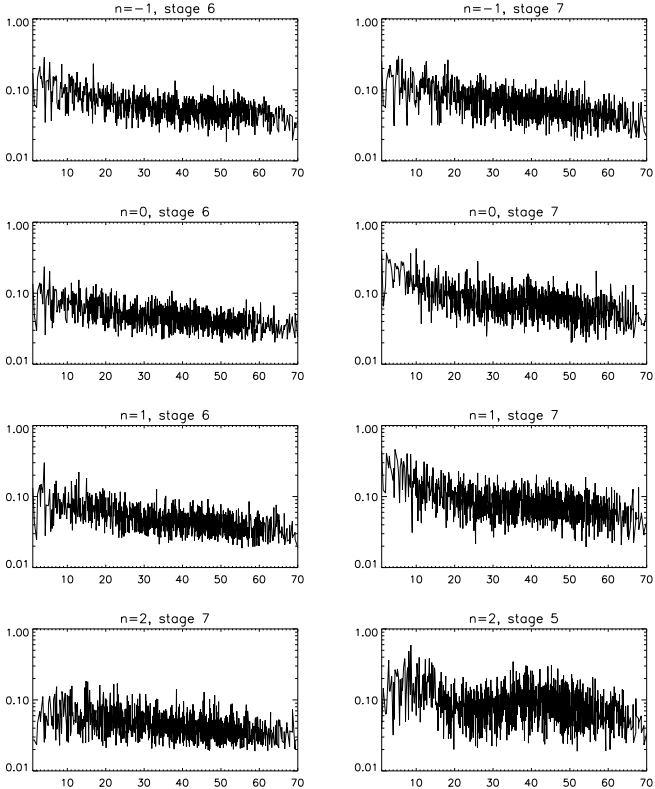,width=9cm} \caption{The
maximal $\meanchi$ is plotted against 1D Fourier scales $|\Delta{\bi
k}| \equiv \sqrt{m^2+n^2}$ for stage 6 and 7 of simulations from
initial spectral index $n=-1$, 0, 1 and 2. The maxima are taken from
each $|\Delta{\bi k}|$ ring of the $(m,n)$ plane. Note that the scale
of $y$-axis is 10 times larger than that in Fig.~\ref{maxchi23} and
\ref{maxchi45}.} \label{maxchi67}
\end{figure}

In Fig.~\ref{maxchi23}, \ref{maxchi45} and \ref{maxchi67} we plot the
maximum $\meanchi$ against $|\Delta {\bi k}|$ for all 4 sets of
simulations. These 1D plots show that the maximal phase coupling
from each ring $|\Delta {\bi k}| \equiv \sqrt{m^2+n^2}$ of the $(m,n)$
plane. As we control the simulations by assigning the same set of
initial random phases, These 1D plots will display how the scale
$|\Delta {\bi k}|$ of phase coupling is related to morphology.

The $N$-body simulations we have carried out is of self-similar nature, that
is, a distribution function $f({\bi x},t)$ has the same statistical
measure as the re-scaled one
\begin{equation}
f \mapsto \lambda^{\alpha} f({\bi x}/\lambda^{\beta},\lambda t),
\mbox{ as $t \mapsto \lambda t$}.
\end{equation}
With the reciprocal-scaling property of Fourier transform,
\begin{equation}
f(\alpha {\bi x}) = \frac{1}{|\alpha|} F\left(\frac{\bi k}{\alpha}\right),
\end{equation}
for non-linear scales ${\bi x}_{\rm NL}$ increasing, i.e. $\alpha
> 1$, the corresponding scales in $k$ space are decreasing.
Although phases do not possess a linear relationship owing to
their circular nature, it can be understood qualitatively that, if
there exists coupling between pairs of phases with fixed $\Delta
{\bi k}$, this scale has to decrease as gravitational clustering
proceeds.

It is therefore clear that for the case such as $n=-1$, where
large-scale filaments are the prominent feature, phase coupling
starts from low $\Delta {\bi k}$, which also has cascade effect on
to higher $\Delta {\bi k}$, as phases strongly couple on any
$\Delta {\bi k}$ might also do so at multiples of $\Delta {\bi
k}$. For high $n$, on the other hand, where small clumps form
first, phase coupling starts from large $\Delta {\bi k}$. This
also explains why the Shannon entropy from neighbouring phase
difference can produce considerable results at early stages for
$n=-1$, but not for $n=2$ (see the fig.5 of Chiang \& Coles 2000).
On the other hand, the coupling between the amplitudes are
enhanced by the factor $1/\alpha$ as clustering proceeds, so
mode-mode coupling between amplitudes at early stages is not as
obvious as between phases. For a discussion of how this relates to
the development of phase correlations in mildly non-linear
evolution, see Watts \& Coles (2002).

The $\meanchi$ statistics shown on the $(m,n)$ plane and those 1D
plots confirm the visualization of phase coupling presented by Coles
\& Chiang (2000). Phase coupling firstly appears on large $\Delta
{\bi k}$ when the the scale of non-linearity is small in real space, then it
shifts on the $(m,n)$ plane to small $\Delta {\bi k}$ in
Fig.~\ref{mnplane1} (c) and (d) and finally dominates at neighbouring
modes as seen in Fig.~\ref{maxchi67}.

\section{Conclusion}
We have generalized a method based on phase mapping on the return
map. This simple, easy-to-implement method can detect phase
coupling at any scales $\Delta {\bi k}$ in  $k$ space.  We
apply this method to two-dimensional simulations of gravitational
clustering and the result has shown that even when the evolution
is in the mild non-linear regime, phase coupling on certain scale
is revealed through the $\meanchi$ statistics on the $(m,n)$
plane.

In contrast to other methods, such as the Shannon entropy of the
distribution of neighbouring phase differences \cite{cc1},
this method does not require large number of phases. Moreover, this
approach can detect the {\it scale} of phase coupling through the
phase mapping as shown in Fig.~\ref{mapping}.

With the systematic $N$-body simulations shown in
Fig.~\ref{simulation}, we have also demonstrated in
Fig.~\ref{maxchi23}-\ref{maxchi67} that the scale of phase coupling
differs according to the clustering morphology: modes between {\it small}
$\Delta{\bi k}$ for large-scale filaments, {\it large} for small clumps.

This method reveals a signature of non-linear gravitational
instability, but also offers the opportunity to provide a general
test of Gaussianity that could be applied to cosmic microwave
background temperature maps. In future work we shall evaluate the
effectiveness for such method.

\section*{Acknowledgments}
This paper was supported in part by Danmarks Grundforskningsfond
through its support for the establishment of the Theoretical
Astrophysics Center and by grants RFBR 17625. PC acknowledges support
from PPARC.

\end{document}